# Superoutbursts and grazing eclipses in the dwarf nova V1227 Herculis

Jeremy Shears, Ian Miller, Roger Pickard & Richard Sabo

## Abstract

We present photometry obtained during the 2012 May and September outbursts of the frequently outbursting dwarf nova, V1227 Her. Superhumps were present in both cases with a peak-to peak amplitude of up to 0.28 mag, showing these events to be superoutbursts. We show for the first time that the system undergoes small eclipses with a depth of up to 0.08 mag, lasting 11 to 14 min, which are likely to be grazing eclipses of the accretion disc. The September outburst was the better observed of the two and lasted at least 14 days with an outburst amplitude of approximately 4 magnitudes. The mean superhump period was $P_{sh}$ = 0.065103(20) d. Analysis of eclipse times of minimum gave an orbital period $P_{orb}$ = 0.064419(26) d, although there is some ambiguity due to the relatively short time over which the eclipses were observed. The fractional superhump period excess, $\varepsilon$, was 0.0106(7).

## Introduction

V1227 Her was identified as a dwarf nova by Szkody *et al.* (1), SDSS J165359.06+201010.4, during their search for cataclysmic variables in data from the Sloan Digital Sky Survey (SDSS). Their follow-up photometric observations occurred at a time when the system was in outburst and revealed a periodic hump feature superposed on a declining trend, which they interpreted as superhumps during the decline from an outburst. The superhump period of the three observed humps was 0.0658 d. These results are consistent with it being a dwarf nova of the SU UMa class. Kato *et al.* (2) performed photometry during an outburst in 2010 May which also revealed superhumps with $P_{sh}$ = 0.065032 d.

Table 1 lists 16 outbursts of V1227 Her, in which the object became brighter that magnitude 16.5, using data from Catalina Real-Time Transient Survey (CRTS) (3), the AAVSO International Database and the present authors. Four of the outbursts were confirmed as superoutbursts; others might well have been, but no time series photometry was conducted to check for superhumps. The short interval between several of the outbursts, as short as 40 days, suggests that V1227 Her is a frequently outbursting dwarf nova. The interval between the 2012 May and September superoutbursts points to a superoutburst period of around 120 d. In this paper we report photometry during these two superoutbursts.

## Photometry

We performed 18 h of photometry during the 2012 May outburst of V1227 Her and 31 h during the September outburst. Our instrumentation is shown in Table 2 and the observation log in Table 3. All but one of the photometry runs were 3.3 hrs or less due to the rather short nights in May at our temperate latitudes and the object's poor location in the western sky in September. Images were dark-subtracted and flat-fielded prior to being measured using differential aperture photometry relative to the BAA VSS V-band sequence P100617 (4).

## The 2012 September outburst

The outburst light curve is shown in Figure 1a. The outburst was detected on Sep 10.986 by JS (5), using the 35 cm Bradford Robotic Telescope, when the star was still brightening. It reached maximum magnitude of 14.6 two days later and faded at a rate of 0.6 mag/d over the next two days to mag 15.0. This marked the beginning of the plateau phase. There was then a brief rebrightening to mag 14.8, followed by a gradual decline at 0.1 mag/ d for the rest of the plateau phase, which lasted about 7 days. Some 14 days after the outburst was detected, the rapid decline set in, although the approach to quiescence itself was missed. SDSS lists the system at g=18.66 in quiescence. Thus the outburst amplitude was approximately 4 mag and its duration was at least 14 days.

Figure 2 shows expanded plots of some of the longer photometry runs. Superhumps were present throughout the outburst, confirming it to be a superoutburst. We measured the times of maximum of 18 superhumps, and their uncertainties, using the Kwee and van Woerden method (6) in the *Minima* v2.3 software (7). These are listed in Table 4. Following a preliminary assignment of superhump cycle numbers to these maxima, an unweighted linear analysis of the times of maximum between HJD 2456182 and 2456189 allowed us to obtain the following superhump maximum ephemeris:

$$HJD_{max} = 2456182.6755(11) + 0.065103(20) \times E \quad \text{[Eqn 1]}$$

Thus the mean superhump period in this interval was $P_{sh} = 0.065103(20)$ d. The observed minus calculated (O–C) residuals for all the superhump maxima relative to the ephemeris are shown in Figure 1b. The downward trend at the beginning of the O-C diagram suggests that the superhump period was very slightly shorter during the first part of the outburst: analysing the times of maximum between HJD 2456182 and 2456185 gave $P_{sh} = 0.064905(31)$ d.

The superhumps increased in size up to a maximum peak-to-peak amplitude of 0.28 mag which corresponded to the time of maximum brightness. They then gradually reduced in amplitude during the first part of the plateau corresponding to the more rapid fade which was described above. We note that the interval over which this occurred also corresponds to the slightly shorter superhump regime. After this point the superhump amplitude remained at 0.11 to 0.15 mag for the rest of the outburst.

Close inspection of the light curve revealed the presence of several small V-shaped dips superimposed on the superhumps, examples of which are shown in Figure 3. The shape and regularity of these leads us to interpret them as eclipses. The eclipses were only obvious when they coincided with superhump minima. We measured 8 times of eclipse minimum, again using the Kwee and van Woerden method in the *Minima* software, and these are shown in Table 5. We obtained the following eclipse ephemeris from an unweighted analysis of these times of minimum:

$$HJD_{min} = 2456183.3720(19) + 0.064419(26) \times E \quad \text{[Eqn 2]}$$

Taking the eclipses as a measure of the orbital period means that $P_{orb} = 0.064419(26)$ d. An O-C diagram relative to this ephemeris is shown in Figure 1d.

The depth and duration of the eclipses was difficult to determine accurately due to the difficulty of isolating the eclipses from the superhump profile. The eclipse depth varied between 0.03 and 0.08 mag with a mean of 0.06 mag (Table 5) and the eclipse duration varied between 11 and 14 min.

**2012 May outburst**

The 2012 May outburst was detected by IM on May 14.934 (8), but was much less well observed than the September event. The light curve is shown in Figure 4a. It appears to have been caught near maximum brightness, at mag 14.7, and part of the plateau phase was followed over an interval of eight days. The mean rate of fading during the plateau was 0.11 mag/d, the same as was found in the 2012 September outburst. Superhumps were present (Figure 5) confirming that this was also a superoutburst. Times of 8 superhump maximum were measured as previously and are shown in Table 4.

An unweighted analysis of the times of superhump maximum gave:

$$\text{HJD}_{max} = 2456063.4327(4) + 0.065114(8) \times E \qquad\qquad \text{[Eqn 3]}$$

Thus the superhump period, $P_{sh} = 0.065114(8)$ d, was consistent with the one measured in 2012 September. The O-C diagram (Figure 4b), shows that the superhump period was constant during the observed part of the outburst. The superhump amplitude declined from 0.22 mag to 0.09 mag (Figure 4c and Table 4).

We carefully examined the light curves from this outburst for the presence of eclipses, but could be certain of only two such events (Table 5), with a depth of approximately 0.08 mag. Extending the 2012 September eclipse ephemeris back to May, using all the eclipse timings, gives the following eclipse ephemeris:

$$\text{HJD}_{min} = 2456069.4600(20) + 0.0644295(12) \times E \qquad\qquad \text{[Eqn 4]}$$

Given the long gap between the two sets of eclipses, we considered the change in period that would be caused by a difference of one cycle count during this period:

+1 cycle different

$$\text{HJD}_{min} = 2456069.4598(22) + 0.0643944(13) \times E \qquad\qquad \text{[Eqn 5]}$$

-1 cycle different

$$\text{HJD}_{min} = 2456069.4601(25) + 0.0644647(15) \times E \qquad\qquad \text{[Eqn 6]}$$

The difference in period is only about 1.5 times the error on the period in Eqn 2, so there is a strong possibility that our cycle count could be out by +/- 1. There also remains a remote possibility that there is a 2 cycle error in the count. Clearly timings of eclipses during future outbursts would help to confirm the period and refine the ephemeris.

**The nature of the eclipses**

The short duration and shallowness of the eclipses suggests that these are grazing eclipses of the accretion disc, with the binary system being only slightly above the critical inclination for eclipses to occur. The fact that the eclipses were only evident at some stages of the superoutburst is also consistent with a marginal graze when the accretion disc is sufficiently expanded. The eclipse depth in eclipsing SU UMa systems is often strongly affected by the location relative to the superhumps: eclipses are shallower when hump maximum coincides with eclipse, such as is found, for example, in the case of DV UMa (9).

Close examination of Szkody *et al.*'s light curve of V1227 Her, which appears as Figure 6 of their paper (1), shows inflections in the minima of the superhump light curves which might be indicative of very shallow eclipses. Moreover, there is a suggestion of a third event at the end of their photometry. They did not draw attention to these events and our suggestion that they are eclipses is highly tentative.

The emission spectra of high inclination dwarf novae often show double-peaked and broadened emission lines due to the Doppler shift associated with the rotation of the accretion disc. However, the emission spectrum of V1227 Her, shown in Figure 6, appears to be single-peaked and there is no obvious peak broadening. Although the absence of these features might be due to the inclination being close to the critical value, this is not diagnostic as some high inclination systems have single-peaked spectra. For example, the SU UMa dwarf nova SDSS J081610.84+453010.2 shows 0.4 to 0.6 mag eclipses, but has a single-peaked emission spectrum (10).

**Superhump period excess**

Taking our measured values of $P_{orb}$ = 0.064419(26) d and $P_{sh}$ = 0.065103(20) d from the 2012 September superoutburst, allows the fractional superhump period excess $\varepsilon$= $(P_{sh}-P_{orb})/P_{orb}$ to be calculated as 0.0106(7). We note that this value of $\varepsilon$ is rather small compared with the values observed in other SU UMa dwarf novae with similar $P_{orb}$, which are typically in the range 0.020 to 0.036 (11).

Patterson *et al.* (11) developed an empirical relationship between $\varepsilon$ and the mass ratio of the secondary to the white dwarf primary, q= $M_{sec}/M_{wd}$, in a range of cataclysmic variables. Measuring $\varepsilon$ provides a way of estimating the mass ratio, q= $M_{sec}/M_{wd}$, of a cataclysmic variable and using this relationship we find q = 0.054. Such value of q is at the lower end of the range of q for the cataclysmic variables listed in Patterson *et al.* (11) and suggests either an improbably low-mass secondary or a relatively high mass primary.

**Conclusions**

Our photometry of the 2012 September and May outbursts of V1227 Her shows the presence of superhumps with a peak-to peak amplitude of up to 0.28 mag, confirming these events to be superoutbursts of an SU UMa type dwarf nova. The September outburst was the better observed of the two and lasted at least 14 days with an outburst amplitude of approximately 4 magnitudes. The mean superhump period was $P_{sh}$ = 0.065103(20) d. We show for the first time that the system undergoes small eclipses with a depth of up to 0.08 and lasting 11 to 14 min. These are likely to be grazing eclipses of the accretion disc. Analysis of eclipse times of minimum gave an orbital period of $P_{orb}$ = 0.064419(26) d. Thus the fractional superhump period excess, $\varepsilon$, was 0.0106(7).

V1227 Her appears to undergo frequent outbursts, at intervals as short as 40 d. We propose further monitoring of the system to confirm its outburst period and the length of the supercycle, which may be around 120 days. Furthermore, measurement of additional eclipse times will help to refine the measurement of $P_{orb}$. It would also be worthwhile to conduct photometry during quiescence to ascertain whether eclipses persist when the accretion disc would be at its minimum diameter. This would require a sizeable telescope given the faintness of the system at quiescence.

## Acknowledgments

The authors gratefully acknowledge the use of observations from the AAVSO International Database contributed by observers worldwide. This research made use of data from the Sloan Digital Sky Survey (SDSS) and the Catalina Real-Time Transient Survey. We also used SIMBAD, operated through the Centre de Données Astronomiques (Strasbourg, France) and the NASA/Smithsonian Astrophysics Data System. JS thanks the Department of Cybernetics at the University of Bradford for the use of the Bradford Robotic Telescope (BRT), located at the Teide Observatory on Tenerife in the Canary Islands, in his monitoring programme of cataclysmic variables – the BRT was used to detect the 2012 September outburst of V1227 Her.Finally we thank our referees, Dr. Chris Lloyd and Dr. David Boyd, for their helpful comments which have improved the paper.

## Addresses

JS: "Pemberton", School Lane, Bunbury, Tarporley, Cheshire, CW6 9NR, UK [bunburyobservatory@hotmail.com]
IM: Furzehill House, Ilston, Swansea, SA2 7LE, UK [furzehillobservatory@hotmail.com]
RP:  3 The Birches, Shobden, Leominster, Herefordshire, HR6 9NG, UK [roger.pickard@sky.com]
RS: 2336 Trailcrest Dr., Bozeman, MT 59718, USA [richard@theglobal.net]

| Date (UT) of detection | JD | Days since start of previous outburst | Maximum mag. | Detection of outburst | Confirmed superoutburst? |
|---|---|---|---|---|---|
| 2004 Aug 24 | 2453241.6 | | 14.9 | Ref (1) | Yes - ref (1) |
| 2005 Jul 16 | 2453567.7 | 326 | 14.9 | CRTS | |
| 2006 Oct 20 | 2454028.6 | 461 | 16.3 | CRTS | |
| 2007 Apr 23 | 2454214.0 | 185 | 15.0 | CRTS | |
| 2008 Mar 23 | 2454549.0 | 335 | 15.0 | CRTS | |
| 2008 Jun 8 | 2454625.9 | 77 | 16.2 | CRTS | |
| 2009 Feb 13 | 2454876.0 | 250 | 14.9 | CRTS | |
| 2009 May 4 | 2454955.8 | 80 | 15.5 | CRTS | |
| 2009 Sep 29 | 2455103.6 | 228 | 15.0 | CRTS | |
| 2010 Apr 15 | 2455302.0 | 198 | 14.5 | CRTS | Yes – ref (2) |
| 2010 Jun 16 | 2455363.8 | 62 | 15.2 | CRTS | |
| 2010 Sep 17 | 2455457.5 | 94 | 15.4 | AAVSO | |
| 2010 Oct 27 | 2455497.4 | 40 | 15.2 | AAVSO | |
| 2011 Jan 28 | 2455590.0 | 93 | 16.4 | CRTS | |
| 2012 May 14 | 2456062.4 | 472 | 14.6 | IM | Yes – this study |
| 2012 Sep 10 | 2456181.5 | 119 | 14.6 | JS | Yes – this study |

Table 1: Outbursts of V1227 Her between 2004 Aug and 2012 October

CRTS = detection by CRTS (3); AAVSO = outburst recorded in the AAVSO International database

| Observer | Telescope | CCD | Filter |
|---|---|---|---|
| Miller | 0.35 m SCT | Starlight Xpress SXVR-H16 | None |
| Pickard | 0.4 m SCT | Starlight Xpress SXVF-H9 | V |
| Sabo | 0.43 m reflector | SBIG STL-1001 | None |
| Shears | 0.28 m SCT | Starlight Xpress SXVF-H9 | None |

Table 2: Equipment used

| Date (UT) | Start time (JD) | Duration (h) | Observer |
|---|---|---|---|
| **_2012 September_** | | | |
| September 12 | 2456182.631 | 3.3 | Sabo |
| September 12 | 2456183.328 | 2.4 | Miller |
| September 13 | 2456183.630 | 3.3 | Sabo |
| September 13 | 2456184.310 | 1.8 | Shears |
| September 14 | 2456184.653 | 3.0 | Sabo |
| September 14 | 2456185.335 | 2.3 | Miller |
| September 17 | 2456187.660 | 3.0 | Sabo |
| September 18 | 2456188.603 | 3.3 | Sabo |
| September 18 | 2456189.327 | 0.3 | Shears |
| September 19 | 2456189.599 | 3.3 | Sabo |
| September 19 | 2456190.299 | 0.7 | Shears |
| September 20 | 2456190.628 | 1.1 | Sabo |
| September 21 | 2456191.618 | 1.1 | Sabo |
| September 22 | 2456193.292 | 0.7 | Shears |
| September 23 | 2456193.614 | 1.6 | Sabo |
| **_2012 May_** | | | |
| May 15 | 2456063.403 | 2.5 | Miller |
| May 15 | 2456063.421 | 2.6 | Pickard |
| May 16 | 2456064.394 | 1.8 | Pickard |
| May 20 | 2456068.476 | 2.1 | Miller |
| May 21 | 2456069.429 | 3.2 | Miller |
| May 22 | 2456070.409 | 4.4 | Pickard |
| May 22 | 2456070.451 | 1.1 | Miller |

Table 3: Log of time-series observations

| Superhump cycle no. | Superhump max (HJD) | Uncertainty (d) | O-C (d) | Superhump amplitude (mag) |
|---|---|---|---|---|
| **2012 September** | | | | |
| 0 | 2456182.6787 | 0.0015 | 0.0032 | 0.20 |
| 1 | 2456182.7427 | 0.0010 | 0.0021 | 0.24 |
| 10 | 2456183.3283 | 0.0015 | 0.0017 | 0.28 |
| 11 | 2456183.3935 | 0.0010 | 0.0018 | 0.27 |
| 15 | 2456183.6531 | 0.0015 | 0.0010 | 0.27 |
| 16 | 2456183.7177 | 0.0005 | 0.0005 | 0.26 |
| 25 | 2456184.3045 | 0.0025 | 0.0014 | 0.22 |
| 26 | 2456184.3656 | 0.0030 | -0.0026 | 0.22 |
| 31 | 2456184.6907 | 0.0015 | -0.0030 | 0.23 |
| 32 | 2456184.7549 | 0.0010 | -0.0039 | 0.22 |
| 41 | 2456185.3396 | 0.0025 | -0.0052 | 0.16 |
| 42 | 2456185.4050 | 0.0030 | -0.0049 | 0.20 |
| 77 | 2456187.6904 | 0.0020 | 0.0019 | 0.11 |
| 78 | 2456187.7556 | 0.0035 | 0.0020 | 0.12 |
| 92 | 2456188.6690 | 0.0010 | 0.0040 | 0.14 |
| 93 | 2456188.7331 | 0.0025 | 0.0030 | 0.13 |
| 107 | 2456189.6414 | 0.0015 | -0.0001 | 0.15 |
| 108 | 2456189.7034 | 0.0025 | -0.0033 | 0.12 |
| **2012 May** | | | | |
| 0 | 2456063.4330 | 0.0010 | 0.0003 | 0.21 |
| 0 | 2456063.4331 | 0.0010 | 0.0004 | 0.22 |
| 1 | 2456063.4978 | 0.0010 | 0.0000 | 0.21 |
| 1 | 2456063.4981 | 0.0020 | 0.0003 | 0.21 |
| 15 | 2456064.4083 | 0.0020 | -0.0011 | 0.20 |
| 78 | 2456068.5118 | 0.0015 | 0.0002 | 0.16 |
| 93 | 2456069.4896 | 0.0015 | 0.0013 | 0.10 |
| 94 | 2456069.5522 | 0.0015 | -0.0012 | 0.09 |

Table 4: Superhump maximum times

| Eclipse no. | Eclipse min (HJD) | Uncertainty (d) | O-C (d) | Eclipse depth (mag) | Eclipse duration FWHM (min) |
|---|---|---|---|---|---|
| **2012 September** | | | | | |
| 0 | 2456183.3693 | 0.0010 | -0.0027 | 0.05 | 11 |
| 5 | 2456183.6937 | 0.0005 | -0.0004 | 0.05 | 11 |
| 6 | 2456183.7630 | 0.0015 | 0.0045 | 0.07 | 13 |
| 82 | 2456188.6490 | 0.0010 | -0.0054 | 0.08 | 14 |
| 83 | 2456188.7182 | 0.0010 | -0.0006 | 0.08 | 14 |
| 97 | 2456189.6228 | 0.0015 | 0.0021 | 0.06 | 12 |
| 98 | 2456189.6865 | 0.0010 | 0.0014 | 0.08 | 11 |
| 113 | 2456190.6524 | 0.0025 | 0.0010 | 0.04 | 11 |
| **2012 May** | | | | | |
| | 2456069.4598 | 0.0045 | | 0.08 | 14 |
| | 2456069.5245 | 0.0025 | | 0.08 | 12 |

Table 5: Eclipse times

FWHM = full width at half-minimum

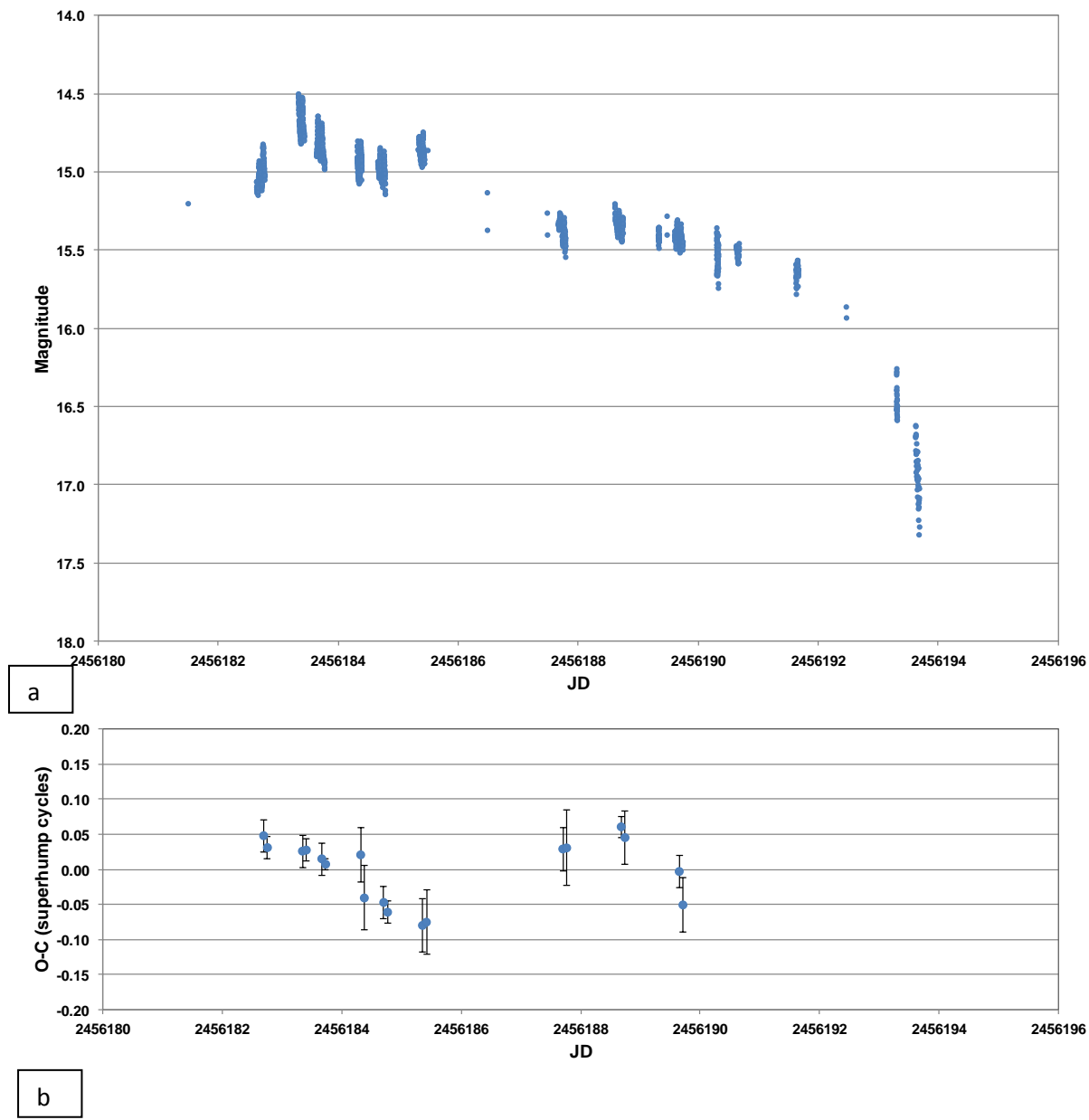

a

b

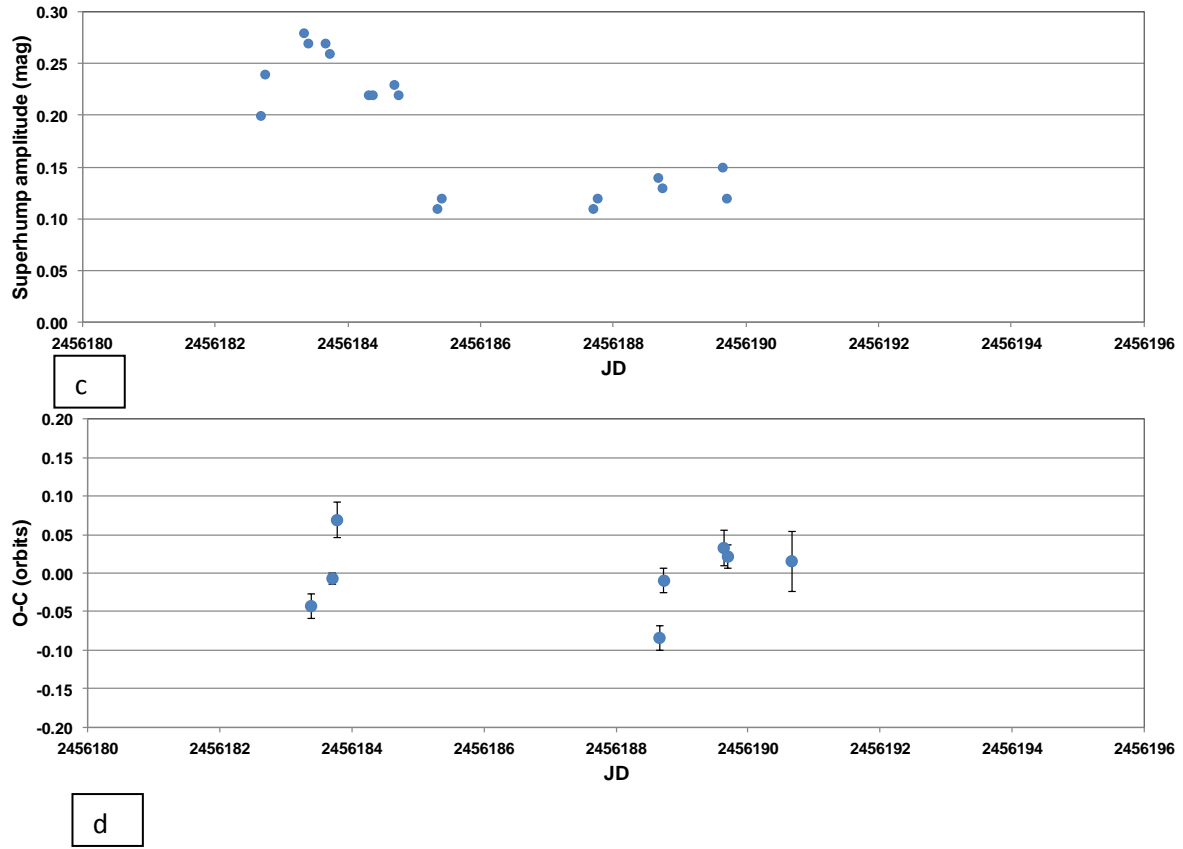

Figure 1 (including previous page): The 2012 September outburst. (a) Outburst light curve. (b) O−C diagram of the superhumps. (c) Superhump amplitude. (d) O-C diagram of the eclipses

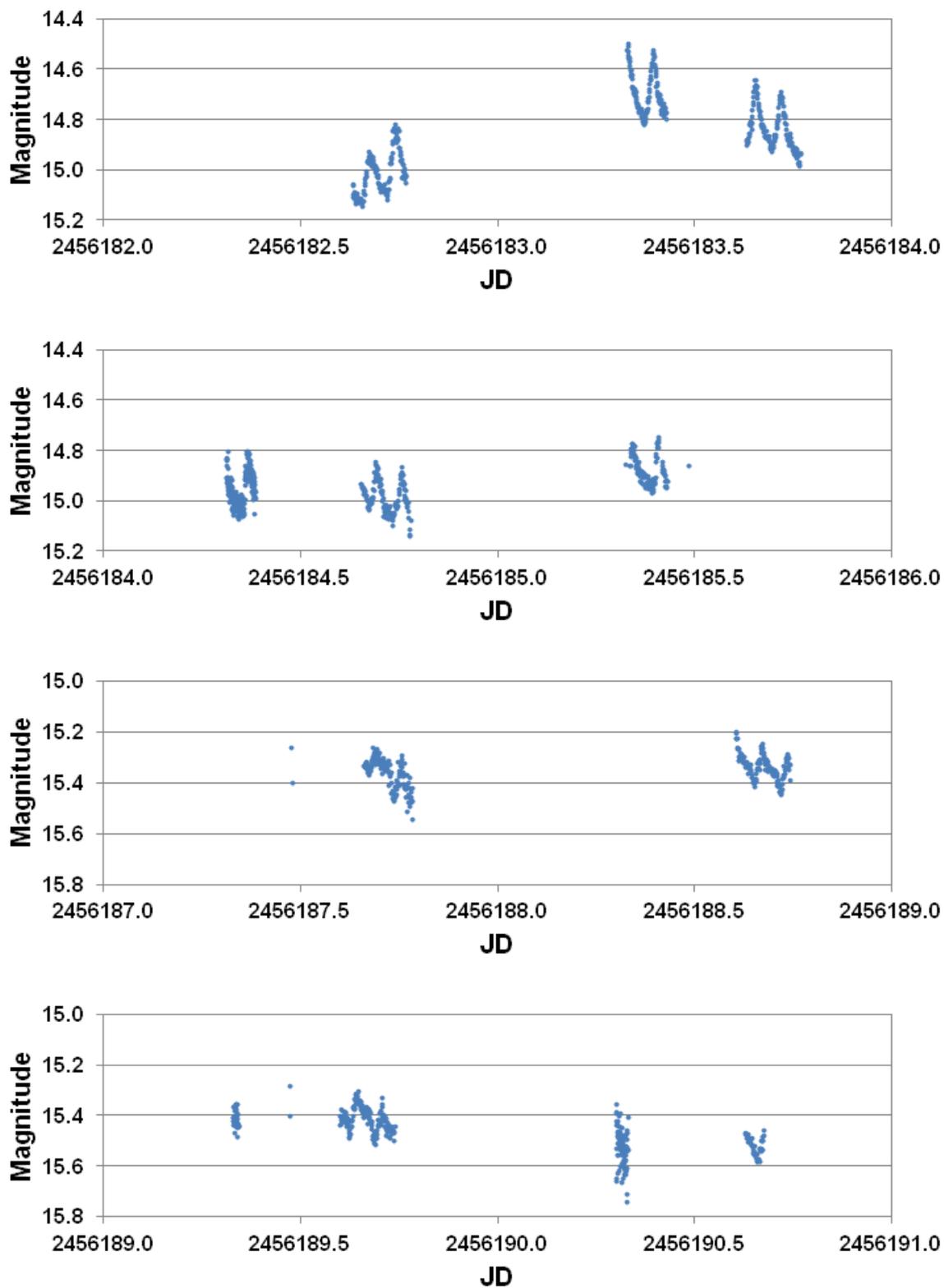

Figure 2: Time resolved photometry during the 2012 September outburst

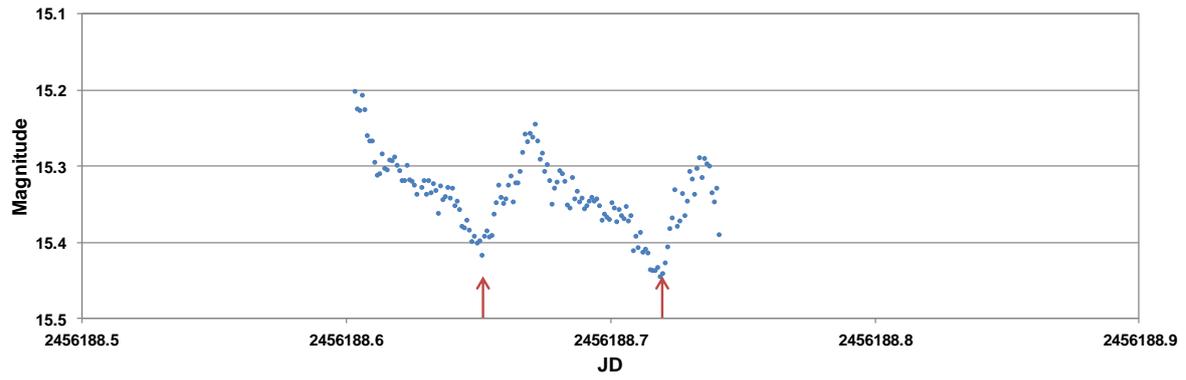

Figure 3: Photometry on HJD 2456188 from the 2012 September outburst showing two eclipses

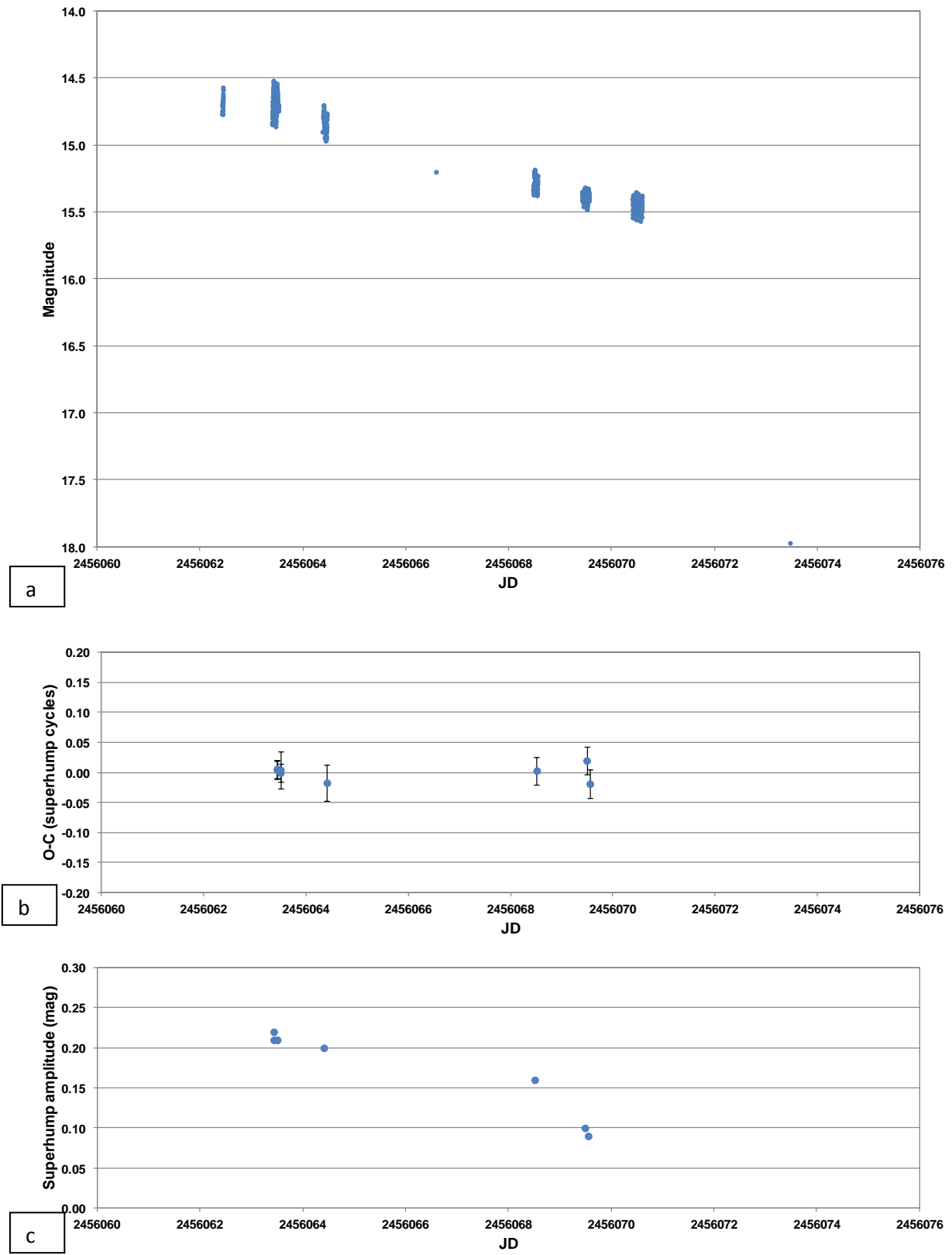

Figure 4: The 2012 May outburst (a) Outburst light curve. (b) O−C diagram of the superhumps. (c) Superhump amplitude.

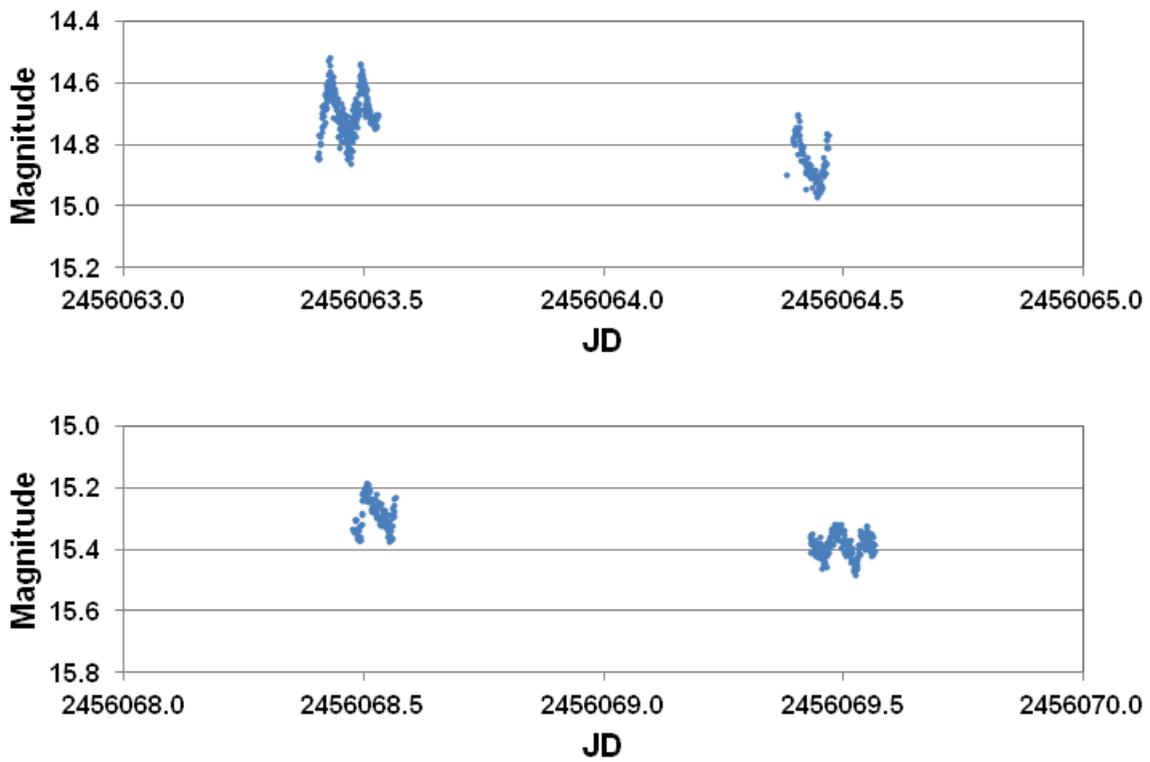

Figure 5: Time resolved photometry during the 2012 May outburst

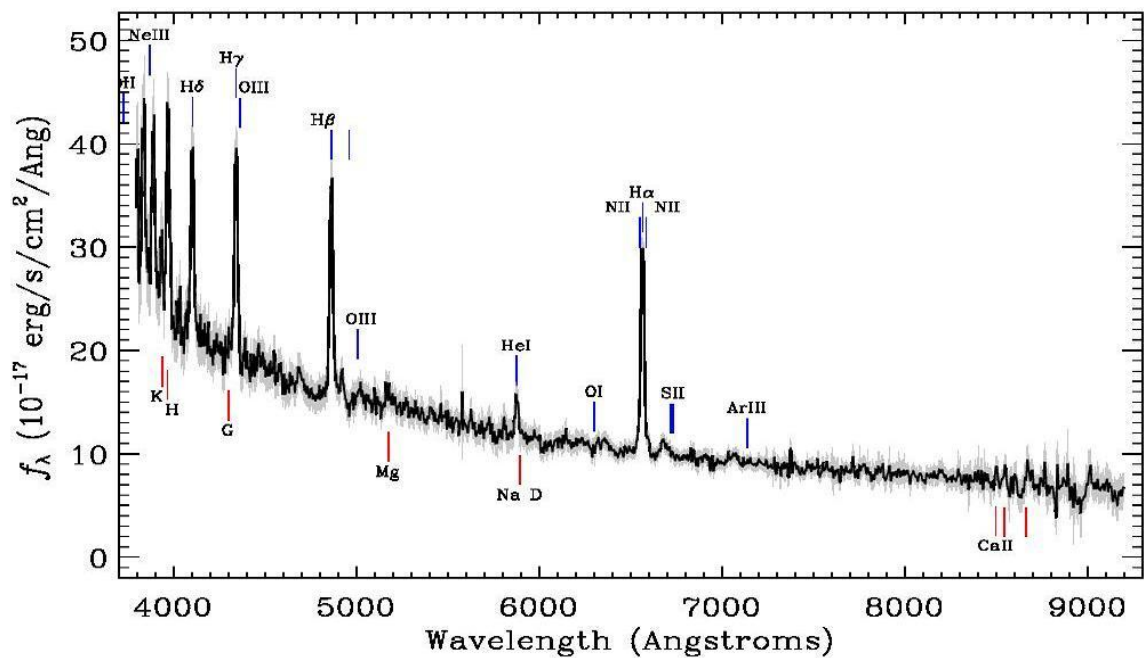

Figure 6: Spectrum of V1227 Her (12)